\documentclass{Interspeech2024}
\usepackage{graphicx}

\usepackage{lipsum}
\usepackage{ragged2e}
\newcommand\blfootnote[1]{%
  \begingroup
  \renewcommand\thefootnote{}\footnote{#1}%
  \addtocounter{footnote}{-1}%
  \endgroup
}
\usepackage{subcaption}

\interspeechcameraready

\title{The Second {DISPLACE} Challenge : DIarization of SPeaker and LAnguage in Conversational Environments}

\name[affiliation={1}]{Shareef Babu}{Kalluri}
\name[affiliation={1}]{Prachi}{Singh}
\name[affiliation={2}]{Pratik Roy} {Chowdhuri}
\name[affiliation={1}]{Apoorva}{Kulkarni}
\name[affiliation={2}]{Shikha} {Baghel}
\name[affiliation={3}]{Pradyoth} {Hegde}
\name[affiliation={3}]{Swapnil} {Sontakke}
\name[affiliation={3}]{Deepak} {K T}
\name[affiliation={4}]{S. R. Mahadeva} {Prasanna}
\name[affiliation={2}]{Deepu} {Vijayasenan}
\name[affiliation={1}]{Sriram} {Ganapathy}
\address{
  $^1$LEAP Lab, Electrical Engineering, IISc, Bangalore,
  $^2$Department of E$\&$C, NITK, Surathkal,
  $^3$IIIT Dharwad,
  $^4$IIT Dharwad, India.
  }
\email{shareefbabu1@gmail.com}


\begin{document}

\maketitle
\begin{abstract}

The DIarization of SPeaker and LAnguage in Conversational Environments (DISPLACE) 2024 challenge is the second in the series of DISPLACE challenges, which involves tasks of speaker diarization (SD) and language diarization (LD) on a challenging multilingual conversational speech dataset. 
In the DISPLACE 2024 challenge, we also introduced the task of automatic speech recognition (ASR) on this dataset. The dataset containing $158$ hours of speech, consisting of both supervised and unsupervised mono-channel far-field recordings, was released for LD and SD tracks. Further, $12$ hours of close-field mono-channel recordings were provided for the ASR track conducted on $5$ Indian languages.
The details of the dataset, baseline systems and the leader board results are highlighted in this paper. 
We have also compared our baseline models and the team's performances on evaluation data of DISPLACE-2023 to emphasize the advancements made in this second version of the challenge.  

\end{abstract}


\section{Introduction}

\blfootnote{This work was funded by the project National Language Translation Mission (NLTM): BHASHINI, the Ministry of Electronics and Information Technology (MeitY), Government of India SP/MITO-22-001 grant, and the British Telecom (BT) grants on conversational speech analytics.}

In multilingual cultures, social interactions frequently comprise code-mixed or code-switched speech \cite{bullock2009cambridge,yilmaz2017language}.\textit{ Code-mixing} occurs when morphemes or words from a secondary language are utilized in a primary language phrase 
(Eg.: `\textit{`Wirst du mitmachen, um mit mir ein IPL-Match anzusehen?}'' (Will you join to watch IPL match with me ?)). 
In contrast, \textit{code-switching} involves modifying the conversational language itself at the sentence or phrase level. (Eg.: 
`` \textit{I'm busy today but,  Ich kann beim nächsten Spiel mitmachen}''
(I'm busy today but, I can join for next match)).
In multilingual communities like Asia, Europe, America and some parts of the African continent, code-mixing and code-switching are more frequent in social conversations \cite{auer2013code,potowski2011bilingual,ganji2019iitg}. 

The code-mixed or code-switched instances pose significant challenges for speech-based systems, such as speaker and language identification or automatic speech recognition (ASR). In multi-speaker and multilingual scenarios, the task of identifying ``\textit{who spoke when}" and ``\textit{which language was spoken when}", termed as speaker diarization (SD) and language diarization (LD) respectively, are significantly challenging. We find that the current speech processing systems are ill-equipped to perform these tasks meaningfully \cite{baghel23_interspeech}. 
 
In this paper, we detail our efforts in extending the first DISPLACE challenge conducted in $2023$ \cite{baghel23_interspeech}. 
We created a dataset for the second \textit{DISPLACE} challenge\footnote{\url{https://displace2024.github.io/}} from four different academic institutes preserving the same recording settings for the entire dataset. The data reflects the social interactions in multilingual communities with code-mixed or code-switched speech, natural overlaps, reverberation, and noise.
The key highlights  of the DISPLACE-2024 challenge are, 
\begin{itemize}
\item Introducing the track on speech recognition which attempts to investigate speech transcription in code-mixed multi-speaker settings on $5$ different languages.
\item Releasing $38$ hours of annotated data and $120$ hours  of unsupervised data. 
\item  Updated baseline systems on speaker and language diarization, which improved the benchmark significantly over the DISPLACE-2023 challenge. 

\item A leader-board platform
\footnote{\url{https://codalab.lisn.upsaclay.fr/competitions/17682}} 
for all $3$ tracks for participants to monitor their progress in system development.  

 \end{itemize}

    
        
  
        
    

\begin{figure*}[htb]
    \centering
    \includegraphics[width=\linewidth]{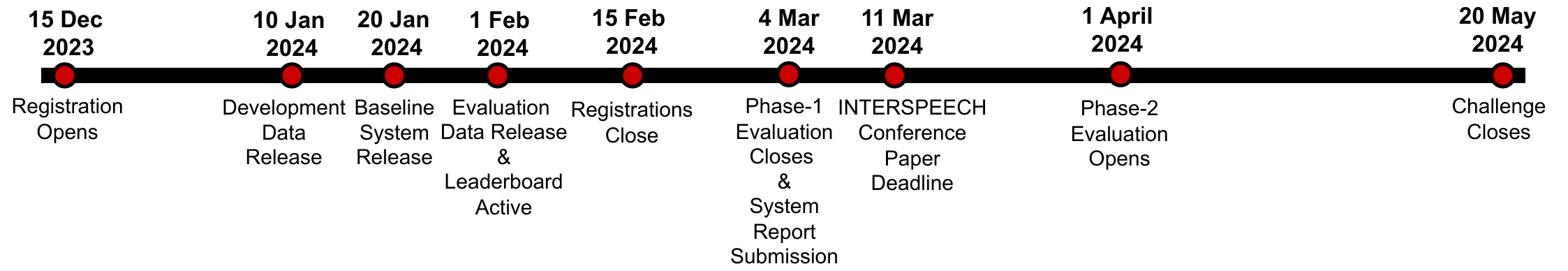}
    \caption{The DISPLACE 2024 Challenge Timeline.}
    \label{fig:timeline}\vspace{-15pt}
\end{figure*}

\section{Related work}

\subsection{Speaker Diarization}
Early research in speaker diarization was predominantly influenced by the evaluations conducted by the National Institute of Standards and Technology-Rich Transcription (NIST-RT) \cite{NIST-RT} on broadcast news (BN) and informal telephone conversations in English. 
The Diarization Error Rate (DER), which continues to be the primary evaluation metric for SD systems, was also proposed during the NIST-RT evaluations.
A series of SD works by Fiscus et al. \cite{fiscus2007rich} used in-domain conversations from meetings. 
Recently, there have been several evaluation challenges on SD, namely, the DIHARD challenge \cite{ryant2020third},  VoxSRC-20 \cite{nagrani2020voxsrc} using YouTube videos and Fearless Steps Series  \cite{joglekar2020fearless} in multi-party and multi-stream naturalistic audio.
\subsection{Language Diarization}
For the past decade, language diarization has been one of the notable research domains in the multilingual speech-processing research community \cite{liu2023reducing}.
Works carried out for LD tasks mainly used broadcast datasets \cite{yilmaz2017language}, and recorded in closed environments \cite{v18_sltu}.
Recently, MERLIon CCS Challenge on language identification and diarization used code-switched child-directed speech \cite{chua23_interspeech}. 

\subsection{Automatic Speech Recognition}
ASR systems are majorly developed for monolingual scenarios in clean speech data and are less efficient on low-resourced speech data \cite{schultz2001language,srivastava18_sltu}. ASR Challenges at Interspeech $2018$ have targeted read-out and conversational speech data in low-resource Indian languages like Tamil, Telugu, and Gujarati \cite{srivastava18_sltu} and scripted telephone conversations of Assamese, Bengali, Tamil, Urdu, and Hindi \cite{SR_indianLang}. A series of ASR challenges \cite{IITM_HIn_ASR, IITM_hiTaeng_ASR},  on Tamil, Hindi, and Indian-English targeted lecture data (formal style) and read speech.  But code-switching scenarios were explored in limited languages like Bengali-English, Hindi-English \cite{diwan2021multilingual}. 
There have been efforts in non-Indian languages, such as English and Arabic, to tackle speech recognition in code-switching \cite{MCSR} and the Oriental Language Recognition challenge for multilingual societies \cite{li2021oriental}.


Despite several recent efforts to provide real labeled conversational datasets through various challenges \cite{ryant2020third,joglekar2020fearless,nagrani2020voxsrc,chua23_interspeech}, there is still a need for a more generic dataset that can capture multi-speaker, multilingual, code-switched scenarios that are common in day-to-day conversations. 
DISPLACE 2024 challenge tries to address this need by providing a natural multilingual conversational dataset without any restrictions on speakers, languages, topics, etc. The challenge evaluates the SD, LD, and ASR performance on the same dataset to make it more meaningful. 

\section{{DISPLACE} Corpus Details }
{DISPLACE} corpus has been collected from four different academic institutions. Each institute followed  the same recording protocol and instrument specifications but differed  in the recording room settings like size, shape, and acoustic properties. 
The recorded conversations contained different speakers across different age groups and different regions of India.
The procedure in participant selection ensured their ability to converse fluently with each other in one of the Indian languages (L1) along with Indian-English. 
Each recording session lasted about $30$-$60$ minutes, with $3$-$5$ participants conversing in their native language, along with natural code-mixing and code-switching instances of Indian English. 
The topics for the conversations were mainly about culture, lifestyles, entertainment, and sports, excluding  emotionally sensitive or personal topics. 
The data is collected using lapel microphones and omnidirectional microphone recorders for close-field and far-field recordings respectively.  More details of the recording setup and participant selection are given in \cite{baghel23_interspeech, baghel2023summary}. 

Professional annotators were hired to label the data using close-field recordings for all the targeted speakers in a session. This annotation process involves marking the details of the speaker, language, and transcript of speech region. A single Rich Transcription Time Marked (RTTM) annotation file is generated for each conversation by combining the annotations obtained from all the participants' lapel microphones. Multiple quality checks were performed on the annotations before we released them to the participants. More details on the data annotation and labeling can be found in \cite{baghel2023summary}.
\subsection{Development and Evaluation set }
\label{subsec:data}

In {DISPLACE} 2024 challenge, we have released development and evaluation data to the participants through {Zenodo} (a cloud service) with a password-protected link\footnote{\url{https://zenodo.org/records/10669296}}. As was done for the DISPLACE-2023 challenge, no training data was provided to the participants, and they were allowed to use any public data resources and /or proprietary data to train their systems.  

In this challenge, we have released both supervised (labeled) and unsupervised (unlabeled) data containing $666$ unique speakers, out of which $459$ are male and $207$ are female speakers.
All the speakers fall in the age group of $17$-$65$ years.
There are $9$ different Indian languages spoken in the $237$ sessions, along with Indian English. 
For the second {DISPLACE} challenge, we have combined the previous challenge's annotated data with $20$\% of additional annotated data.
In total, we have annotated data of $38$ hours of conversational speech from $67$ sessions with $197$ speakers.
The data is pre-processed for volume normalization, all the close-field recordings are time-aligned with the far-field recordings, and the audio is resampled to $16$~kHz and normalized to the [-1,1] range.

\noindent\textbf{\textit{Development set:}} We released far-field supervised (labeled) and unsupervised (unlabeled) conversational data of around $140$ hours as a part of the development (Dev) set to all the registered participants for SD and LD tracks. For the ASR task, we released $4$ hours of supervised data for both far-field and close-field recordings. 
 We have released annotated data of $20$ hours of conversational speech from $35$ sessions with $98$ speakers as a part of supervised Dev data for diarization tracks. The   sessions chosen for speaker and language diarization tracks are identical, and  the labels are provided separately for each track in RTTM format. 
We have released supervised data for the ASR track for four Indian languages (i.e., Bengali, Hindi, Kannada, Telugu) and Indian-accented English. Each native language has one hour of labeled transcripts for close and far-field recordings. For the ASR track, we provided the segment files to identify the speech regions along with the transcript labels in native language script.   

\raggedright{ \textbf{\textit{Unsupervised Data:}} }
\justifying We   released unsupervised data   in the second  {DISPLACE} challenge. 
In total, we distributed more than $120$ hours of conversational speech data from $170$ sessions with $493$ speakers from Indian languages like Hindi, Bengali, Kannada, Telugu, Malayalam, Tamil, Marathi, Assamese, Odiya and Indian English. This unsupervised data was meant to be useful for adapting the models to the environmental conditions of the DISPLACE evaluation data for all three tracks. 

\raggedright{ \textbf{\textit{Evaluation set:}} }
\justifying As a part of the evaluation (Eval) set, we released the supervised conversational data of $18$ hours from $99$ speakers in $32$ sessions. We use the same far-field evaluation data for SD and LD tasks. For the ASR task, we have released $8$ hours of the close-field recordings.

\section{Challenge Tasks and Organisation}
Participants were encouraged to build their own speech activity detection systems. The submissions are evaluated based on speech regions ignoring non-speech regions like background speech, noise, laughing, or clapping.
The second DISPLACE challenge had the following tracks,

\begin{itemize}
    \item \textbf{\textit{Track-1:}} Speaker Diarization in multilingual scenarios.
    \item \textbf{\textit{Track-2:}} Language Diarization in multi-speaker settings.
    \item \textbf{\textit{Track-3:}} ASR  on single-speaker code-mixed settings.
\end{itemize}

The metric for evaluating the system performance of diarization systems is the diarization error rate (DER) while word error rate (WER) is used for evaluating speech recognition systems \cite{sclite}. 
The challenge's timeline is shown in Fig~\ref{fig:timeline}.

\section{ Baseline Systems}
\subsection{Speaker Diarization}
The   speaker diarization system follows the steps outlined in the DISPLACE 2023 challenge baseline \cite{baghel23_interspeech}, which includes speech activity detection, segmenting audio into $1.5$s chunks with a $250$ms shift, followed by extracting x-vectors using a pre-trained $13$-layer ETDNN model \cite{snyder2019speaker}. These x-vectors are utilized to generate PLDA similarity scores for spectral clustering. The number of speakers is decided based on the threshold set according to the best DER on the dev set. We perform Variational Bayes (VB) re-segmentation \cite{singh2019leap} using the clustering results to generate the final output. Compared to the previous year's baseline, we have integrated Pyannote speech activity detection \cite{bredin23_interspeech} and overlap detection modules. Based on the posterior probabilities from the VB-hidden Markov model (HMM) module and the output of the overlap detection system, we predict up to two speakers for each audio segment. 
    
\subsection{Language Diarization}
The baseline system follows a two-stage methodology, comprising feature extraction and clustering. We used the Pyannote speech activity detection \cite{bredin23_interspeech} for detecting the speech regions.
We then segment the speech regions into short overlapping segments of $10$s with a $200$ms shift. 
Unlike DISPLACE 2023 baseline, which relied on language identification (LID) ECAPA-TDNN model embeddings \cite{baghel23_interspeech}, we employ a deep multitask model called  Whisper \cite{radford2023robust} for the feature extraction. 
The Whisper model is trained on the Voxlingua107 dataset \cite{valk2021slt}, containing data from $99$ languages and $6628$ hours. Notably, the LID system is trained on $30$s speech segments.
Upon feature extraction, the Whisper language detector \cite{vachhani23_interspeech} furnishes $99$ language posterior probabilities. These extracted posterior features serve as input for an agglomerative clustering algorithm. 
We use a cosine similarity-based distance metric computed using the posteriors from the Whisper model for clustering. 
Subsequently, we enhance our results by employing a Variational Bayes x-vector (VBx) model \cite{landini2022bayesian} for further refinement.

We tested our  baseline system with DISPLACE-2023 data and report the development (dev) and evaluation (Eval) DER for both the models (DISPLACE-2023 and the DISPLACE-2024 baseline system) in Table  \ref{tab:LD_baseline}. We observe absolute improvement   of $5.94$\% on dev and $12.11$\% on the Eval data, respectively, in terms of DER for the LD task, whereas it is more than 2\% absolute improvement in DER for SD task on both dev and Eval sets.
\begin{table}[t!]
    \centering
    \scriptsize
    \caption{DER ($\%$) comparison of  LD and SD baseline systems using DISPLACE 2023 and DISPLACE 2024 challenge models on DISPLACE 2023 data.  }
    \renewcommand{\arraystretch}{1.1}
    \begin{tabular}{l|c|c|c|c} \hline \hline
     & \multicolumn{2}{c|}{LD} & \multicolumn{2}{c}{SD} \\ \hline
         Baseline &  Dev &  Eval  &  Dev &  Eval \\ \hline
         {DISPLACE-2023} \cite{baghel2023summary} & 46.95 & 41.67 &27.33 & 32.18\\   
         {DISPLACE-2024} & 41.01 &29.56 & 25.45 & 29.96 \\ \hline \hline 
    \end{tabular}    
    \label{tab:LD_baseline}
    \vspace{-10pt}
\end{table}

\subsection{Automatic Speech Recognition}
As part of the dev and eval phases, we have released  Hindi, Bengali, Kannada, and Telugu conversations. We have provided the segment boundaries and ground truth transcripts for the dev data, while only segment boundary labels were released for the eval data. 
As a part of the baseline system for ASR track, we have implemented the Google Speech-to-Text cloud services using the close field recordings of development data \cite{google_ASR}. 
To compute the WER, we use Sclite toolkit \cite{sclite}, and the WER for dev and eval sets are tabulated in Table \ref{tab:ASR_baseline}. 
As the data was significantly challenging, we observe that the baseline  WER is similar to those observed in other ASR challenges like  CHiME-6  multi-speaker far-field conversational dataset \cite{watanabe2020chime}. Furthermore, the DISPLACE data had code-switched speech, making the ASR task even more demanding.

\begin{table}[t!]
    
    \centering
    \scriptsize
    \renewcommand{\arraystretch}{1.2}
    \caption{Comparison of Track 3- ASR baseline WER with top performing team } 
    \begin{tabular}{l|c|c|c|c} \hline \hline
     & \multicolumn{2}{c|}{Baseline} & \multicolumn{2}{c}{T1} \\ \hline
         Language &  Dev &  Eval  &  Dev &  Eval \\ \hline
        Bengali & 63.5 & 75.23 & 54.79 & 63.87\\  
         Hindi & 58.5  & 60.43 & 38.90 & 52.27\\  
         Kannada & 80.8 & 81.08 & 65.14 & 62.88\\  
         Telugu &  71.2 & 66.92 & 59.35 & 57.87\\  
         Indian-English & 66.5  & 66.76 & 25.53 & 34.56\\ \hline
         Overall & 66.7 & 67.78 & 37.30 & 47.34\\ \hline \hline
    \end{tabular}
    
    \label{tab:ASR_baseline}\vspace{-15pt}
\end{table}
\vspace{-10pt}
\begin{figure}[t!]
    \centering
    \includegraphics[width=0.45\textwidth]{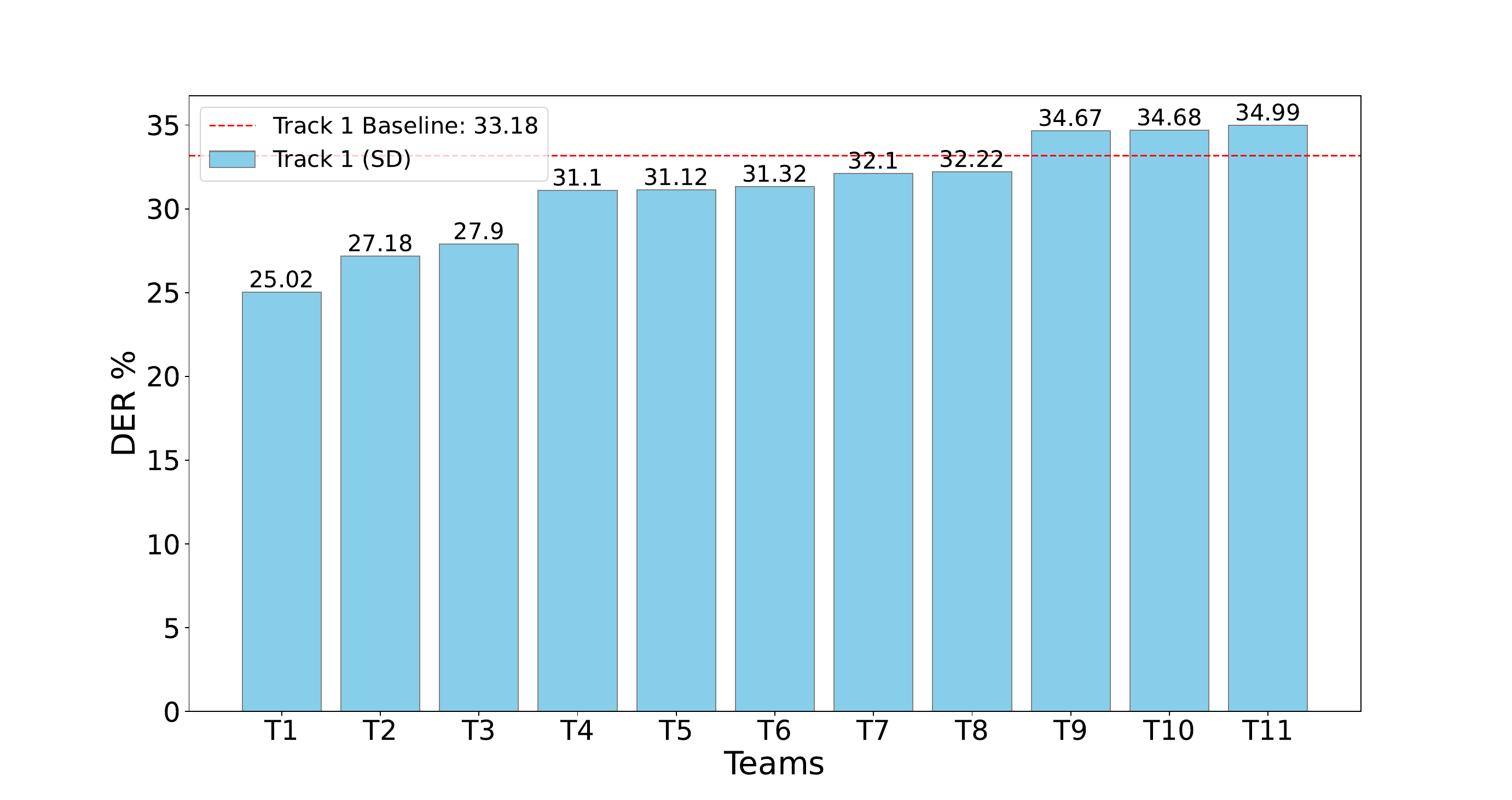}
    \caption{DER ($\%$) comparison of best of Track 1 Eval Phases of Teams submissions with baseline.}
    \label{fig:sd_eval_teams}\vspace{-8pt}
\vspace{-7pt}
\end{figure}
\begin{figure}[t!]
    \centering
    \includegraphics[width=0.45\textwidth]{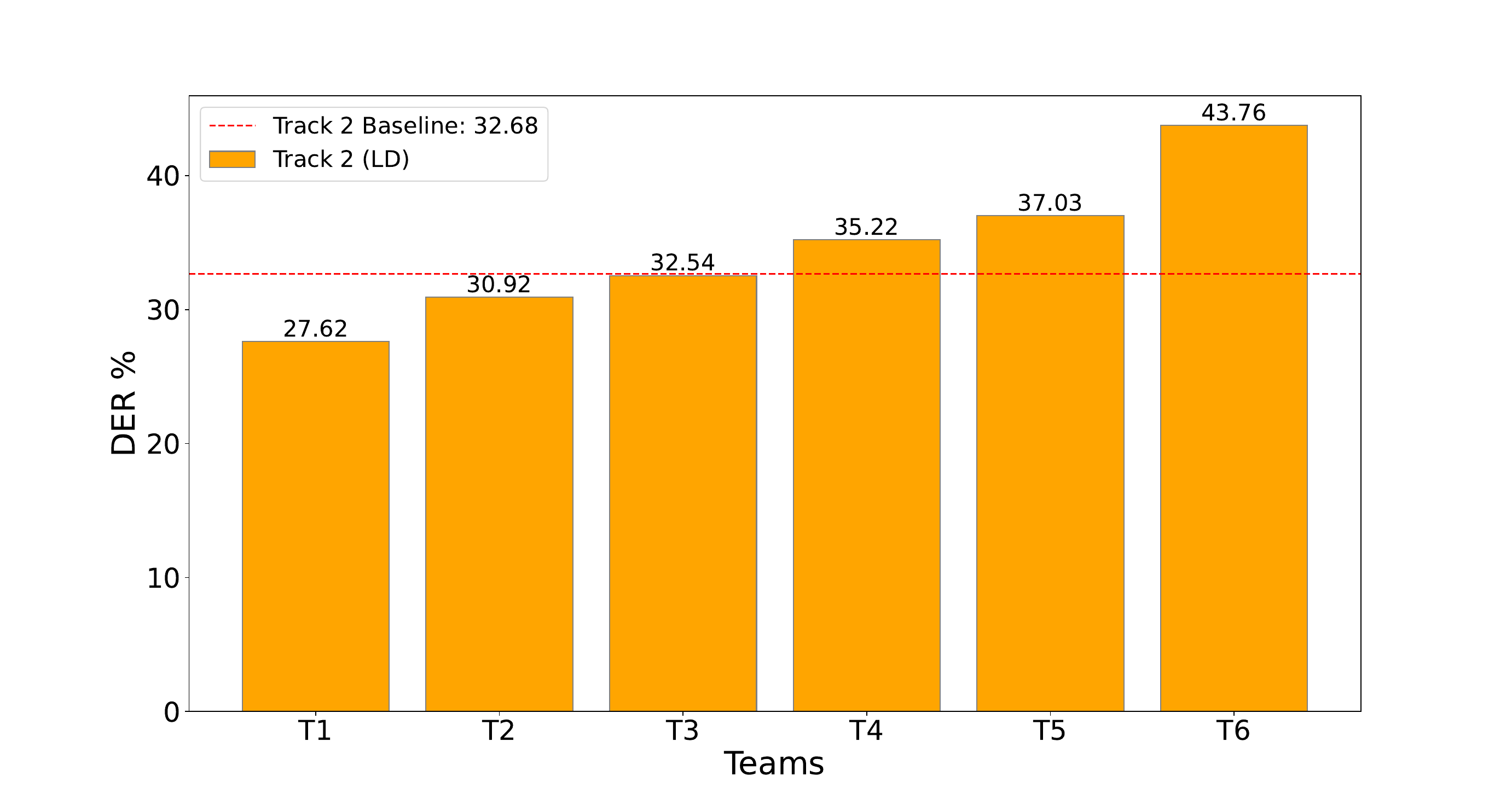}
    \caption{DER ($\%$) comparison of best of Track 2 Eval Phases of Teams submissions with baseline.}
    \label{fig:ld_eval_teams}\vspace{-8pt}
\end{figure}

\begin{table}[]
    \centering
    \scriptsize
    \caption{Comparison of top 3 teams model submissions for SD and LD tracks in terms of DER ($\%$),  DISPLACE-2024 vs DISPLACE 2023 models using DISPLACE-2023 challenge Eval data.}
    \renewcommand{\arraystretch}{1.1}
    \begin{tabular}{l|c|c||c|c} \hline \hline
     & \multicolumn{2}{c||}{Track-1:SD} & \multicolumn{2}{c}{Track-2:LD} \\ \hline
     Team No. & 2024 &  2023  & 2024 &  2023 \\ \hline
        T1 &  21.27 & 27.8 &  25.05 & 37.6\\  
        T2 &  24.61 & 28.6 &  28.57 & 40.2\\  
        T3 &  24.43 & 31.5 &  29.83& 41.2\\ \hline
        \hline
    \end{tabular}
    \label{tab:SD_LD_top3}\vspace{-6pt}
\end{table}

\section{Summary of Challenge Results}
\subsection{Track 1: Speaker Diarization}
A total of $11$ teams participated in  Eval-Phase 1 \& Eval-Phase 2 and eight teams outperformed the baseline. Figure \ref{fig:sd_eval_teams} shows the DER distribution across the teams. 
 We observe a significant boost in the DER performance in \textit{DISPLACE-2024} compared to the best DER obtained in DISPLACE-2023 challenge (absolute improvement of $6.53$\% in DER reported in Table~\ref{tab:SD_LD_top3}).\\
 \noindent\textbf{Top performing systems}: The best single system in T1 submission is the Wav-LM-based speaker segmentation model from \textit{pyannote.audio} \cite{bredin23_interspeech}  fine-tuned on the \textit{DISPLACE-2024} Dev set. The model comprises of end-to-end diarization step followed by global clustering, where the number of speakers is restricted to $7$. 
 The final submission involved an ensemble of Pyannote based models with different configurations,  permutation invariant training (PIT) for speaker diarization and mixture invariant training (MixIT) for speech separation (PixIT) \cite{kalda2024pixit} combined using DOVER-Lap. 
 Team T2 experimented with different pre-trained embedding extractors of the ResNet series \cite{wang2023wespeaker}, and performed spectral clustering and VBx to generate the output. ResNet-152 with spectral clustering achieves the best single system DER. 
 Finally, the submission contained a fusion of the top $7$ models followed by Pyannote overlap detection \cite{bredin23_interspeech}. Team T3 focused on improving voice activity detection (VAD) and overlap detection components. For VAD, the team explored  MarbleNet \cite{jia2021marblenet} and Pyannote SAD \cite{bredin23_interspeech} models   trained using the supervised   and unsupervised   dev data. The best single system comprises multiscale Titanet-L embedding extractor followed by  spectral clustering.

In Table \ref{tab:SD_LD_top3}, we compare the top three teams (T1, T2, and T3) of {DISPLACE-2024} on the same Eval data of DISPLACE-2023. 
This table highlights significant relative improvements for SD with an average relative improvement of $23.49$\% in DER for the top-performing teams.

\subsection{Track 2: Language Diarization}

There were submissions from $6$ teams out of which $3$ teams outperformed over the baseline. The DER distribution across all teams is shown in Figure \ref{fig:ld_eval_teams}. 

\noindent\textbf{Top performing systems}: T1 system used a wav2vec-BERT model \cite{barrault2023seamless} to extract language embeddings from 5s segments with 1s shift. T1 trained a probabilistic linear discriminant analysis (PLDA) model using the DISPLACE ASR dev data. The model is fine-tuned on NIST LRE and SRE data and the dev set. 
This is followed by PLDA similarity scoring, AHC and  VBx re-segmentation to generate the final output. 
T2 system used ResNet34 \cite{gusev2020deep}and wav2vec 2.0 architectures 
for extracting the language embeddings,   utilized spectral clustering of language embeddings within a sliding window and a heuristic bypass (HBP) method for similarity matrix computation.
T3 system used Pyannote \cite{bredin23_interspeech} for VAD and features were extracted using the pre-trained Whisper model \cite{radford2023robust}.  Finally, agglomerative hierarchical clustering was used to cluster and assign language cluster labels to the segments.

In Table \ref{tab:SD_LD_top3}, we compare the top three teams (T1, T2, and T3)   on the same eval data  subset of DISPLACE-2023. 
We observe the relative improvement in the model for LD is around $33.37$\% DER for the top-performing team.

\subsection{Track 3: ASR}
For Track 3, we have two valid submissions from the participants. The top-performing team (T1)    fine-tuned their model using Whisper \cite{radford2023robust} for  English, Hindi, and Bengali languages. The Bhashini model \cite{lodagala2023ccc} was deployed for Kannada and Telugu inferences. This system showed an absolute improvement of $20.4$\% in WER over the baseline system for close-field recordings. The individual  monolingual WER performance along with overall system performance is given Table \ref{tab:ASR_baseline}.

\section{Summary}

This paper provides a comprehensive overview of the second {DISPLACE} challenge, which aims to encourage research in processing multi-lingual multi-speaker conversational audio.  This challenge contained three   tracks: 1) speaker diarization, 2) language diarization and 3) speech recognition. Track-1 and Track-2 share the common data, while Track-3 contained  $12$ hours of audio data in five different Indian languages. As part of the challenge, we released updated baseline systems for the SD and LD tracks which provided improved performance over the first DISPLACE challenge benchmarks. The wide participation for the second DISPLACE challenge across the globe,   resulted in significant improvements over the baseline system for both the SD and LD tracks. The ASR track using close-field recordings was observed to be significantly challenging and thus resulted in limited participation. The best-performing system achieved a WER of $47$\% on the evaluation data, highlighting the need for continued efforts on this demanding dataset. We also compared our baseline models and top performing teams with the corresponding results from the DISPLACE-2023 challenge. This comparison led to understand the  progress  made in processing multilingual, multi-speaker conversational data under the DISPLACE challenge series.

\section{Acknowledgements}

\ifinterspeechfinal
\else
     
\fi
The authors would like to thank Kaustubh Kumar and Lokesh Kumar from IIT Dharwad for helping with data collection. Also Michael free and Shakti Srivastava from BT for valuable discussions.   

\ninept 
\bibliographystyle{IEEEtran}
\tiny \bibliography{mybib}

\begin{thebibliography}{10}
\providecommand{\url}[1]{#1}
\csname url@samestyle\endcsname
\providecommand{\newblock}{\relax}
\providecommand{\bibinfo}[2]{#2}
\providecommand{\BIBentrySTDinterwordspacing}{\spaceskip=0pt\relax}
\providecommand{\BIBentryALTinterwordstretchfactor}{4}
\providecommand{\BIBentryALTinterwordspacing}{\spaceskip=\fontdimen2\font plus
\BIBentryALTinterwordstretchfactor\fontdimen3\font minus
  \fontdimen4\font\relax}
\providecommand{\BIBforeignlanguage}[2]{{%
\expandafter\ifx\csname l@#1\endcsname\relax
\typeout{** WARNING: IEEEtran.bst: No hyphenation pattern has been}%
\typeout{** loaded for the language `#1'. Using the pattern for}%
\typeout{** the default language instead.}%
\else
\language=\csname l@#1\endcsname
\fi
#2}}
\providecommand{\BIBdecl}{\relax}
\BIBdecl

\bibitem{bullock2009cambridge}
B.~E. Bullock and A.~J.~E. Toribio, \emph{The Cambridge handbook of linguistic
  code-switching.}\hskip 1em plus 0.5em minus 0.4em\relax Cambridge University
  Press, 2009.

\bibitem{yilmaz2017language}
E.~Yilmaz, M.~McLaren, H.~van~den Heuvel, and D.~A. van Leeuwen, ``Language
  diarization for semi-supervised bilingual acoustic model training,'' in
  \emph{Proc. IEEE ASRU}, 2017, pp. 91--96.

\bibitem{auer2013code}
P.~Auer, \emph{Code-switching in conversation: Language, Interaction and
  Identity}.\hskip 1em plus 0.5em minus 0.4em\relax Routledge, 2013.

\bibitem{potowski2011bilingual}
K.~Potowski and J.~Rothman, \emph{Bilingual youth: Spanish in English-speaking
  societies}.\hskip 1em plus 0.5em minus 0.4em\relax John Benjamins Publishing,
  2011, vol.~42.

\bibitem{ganji2019iitg}
S.~Ganji, K.~Dhawan, and R.~Sinha, ``{IITG}-{H}ingcos corpus: A {H}inglish
  code-switching database for automatic speech recognition,'' \emph{Speech
  communication}, vol. 110, pp. 76--89, 2019.

\bibitem{baghel23_interspeech}
S.~Baghel, S.~Ramoji, Sidharth, R.~H, P.~Singh, S.~Jain, P.~{Roy Chowdhuri},
  K.~Kulkarni, S.~Padhi, D.~Vijayasenan, and S.~Ganapathy, ``{The DISPLACE
  Challenge 2023 - DIarization of SPeaker and LAnguage in Conversational
  Environments},'' in \emph{Proc. INTERSPEECH 2023}, 2023, pp. 3562--3566.

\bibitem{NIST-RT}
``Rich transcription evaluation,''
  \url{https://www.nist.gov/itl/iad/mig/rich-transcription-evaluation},
  accessed: 2024-02-29.

\bibitem{fiscus2007rich}
J.~G. Fiscus, J.~Ajot, and J.~S. Garofolo, ``The rich transcription 2007
  meeting recognition evaluation,'' in \emph{International Evaluation Workshop
  on Rich Transcription}.\hskip 1em plus 0.5em minus 0.4em\relax Springer,
  2007, pp. 373--389.

\bibitem{ryant2020third}
N.~Ryant, P.~Singh, V.~Krishnamohan, R.~Varma, K.~Church, C.~Cieri, J.~Du,
  S.~Ganapathy, and M.~Liberman, ``{The Third DIHARD Diarization Challenge},''
  in \emph{Proc. Interspeech 2021}, 2021, pp. 3570--3574.

\bibitem{nagrani2020voxsrc}
A.~Nagrani, J.~S. Chung, J.~Huh, A.~Brown, E.~Coto, W.~Xie, M.~McLaren, D.~A.
  Reynolds, and A.~Zisserman, ``Voxsrc 2020: The second voxceleb speaker
  recognition challenge,'' \emph{arXiv preprint arXiv:2012.06867}, 2020.

\bibitem{joglekar2020fearless}
A.~Joglekar, J.~H. Hansen, M.~C. Shekar, and A.~Sangwan, ``{FEARLESS STEPS
  Challenge (FS-2): Supervised Learning with Massive Naturalistic Apollo
  Data},'' in \emph{Proc. Interspeech 2020}, 2020, pp. 2617--2621.

\bibitem{liu2023reducing}
H.~Liu, H.~Xu, L.~P. Garcia, A.~W. Khong, Y.~He, and S.~Khudanpur, ``Reducing
  language confusion for code-switching speech recognition with token-level
  language diarization,'' in \emph{2023 IEEE ICASSP}.\hskip 1em plus 0.5em
  minus 0.4em\relax IEEE, 2023, pp. 1--5.

\bibitem{v18_sltu}
S.~V, V.~Thenkanidiyoor, and D.~A. D, ``{SVM Based Language Diarization for
  Code-Switched Bilingual Indian Speech Using Bottleneck Features},'' in
  \emph{Proc. SLTU 2018}, 2018, pp. 132--136.

\bibitem{chua23_interspeech}
V.~Y.~H. Chua, H.~Liu, L.~P. Garcia, F.~T. Woon, J.~Wong, X.~Zhang,
  S.~Khudanpur, A.~W.~H. Khong, J.~Dauwels, and S.~J. Styles, ``{MERLIon CCS
  Challenge: A English-Mandarin code-switching child-directed speech corpus for
  language identification and diarization},'' in \emph{Proc. INTERSPEECH 2023},
  2023, pp. 4109--4113.

\bibitem{schultz2001language}
T.~Schultz and A.~Waibel, ``Language-independent and language-adaptive acoustic
  modeling for speech recognition,'' \emph{Speech Communication}, vol.~35, no.
  1-2, pp. 31--51, 2001.

\bibitem{srivastava18_sltu}
B.~M.~L. Srivastava, S.~Sitaram, R.~{Kumar Mehta}, K.~{Doss Mohan}, P.~Matani,
  S.~Satpal, K.~Bali, R.~Srikanth, and N.~Nayak, ``{Interspeech 2018 Low
  Resource Automatic Speech Recognition Challenge for Indian Languages},'' in
  \emph{Proc. SLTU 2018}, 2018, pp. 11--14.

\bibitem{SR_indianLang}
``Interspeech 2018 special session : Speech recognition for {I}ndian
  languages,'' \url{https://sites.google.com/view/interspeech2018-ss1},
  accessed: 2024-02-29.

\bibitem{IITM_HIn_ASR}
``Hindi {ASR} challenge,'' \url{https://sites.google.com/view/asr-challenge},
  accessed: 2024-02-29.

\bibitem{IITM_hiTaeng_ASR}
``Hindi-{T}amil-{E}nglish {ASR} challenge,''
  \url{https://sites.google.com/view/indian-language-asrchallenge/}, accessed:
  2024-02-29.

\bibitem{diwan2021multilingual}
A.~Diwan, R.~Vaideeswaran, S.~Shah, A.~Singh, S.~Raghavan, S.~Khare, V.~Unni,
  S.~Vyas, A.~Rajpuria, C.~Yarra \emph{et~al.}, ``Multilingual and
  code-switching asr challenges for low resource indian languages,''
  \emph{arXiv preprint arXiv:2104.00235}, 2021.

\bibitem{MCSR}
``Multilingual and code-switching speech recognition,''
  \url{https://www.clsp.jhu.edu/multilingual-and-code-switching/}, accessed:
  2024-02-29.

\bibitem{li2021oriental}
J.~Li, B.~Wang, Y.~Zhi, Z.~Li, L.~Li, Q.~Hong, and D.~Wang, ``Oriental language
  recognition (olr) 2020: Summary and analysis,'' \emph{arXiv preprint
  arXiv:2107.05365}, 2021.

\bibitem{baghel2023summary}
S.~Baghel, S.~Ramoji, S.~Jain, P.~R. Chowdhuri, P.~Singh, D.~Vijayasenan, and
  S.~Ganapathy, ``Summary of the displace challenge 2023--diarization of
  speaker and language in conversational environments,'' \emph{arXiv preprint
  arXiv:2311.12564}, 2023.

\bibitem{sclite}
``{SCTK}, the {NIST} scoring toolkit,''
  \url{https://github.com/usnistgov/SCTK/}, accessed: 2024-02-29.

\bibitem{snyder2019speaker}
D.~Snyder, D.~Garcia-Romero, G.~Sell, A.~McCree, D.~Povey, and S.~Khudanpur,
  ``Speaker recognition for multi-speaker conversations using x-vectors,'' in
  \emph{ICASSP 2019-2019 IEEE International conference on acoustics, speech and
  signal processing (ICASSP)}.\hskip 1em plus 0.5em minus 0.4em\relax IEEE,
  2019, pp. 5796--5800.

\bibitem{singh2019leap}
P.~Singh, H.~Vardhan, S.~Ganapathy, and A.~Kanagasundaram, ``{LEAP} diarization
  system for the second {DIHARD} challenge,'' in \emph{Proc INTERSPEECH}, 2019,
  pp. 983--987.

\bibitem{bredin23_interspeech}
H.~Bredin, ``{pyannote.audio 2.1 speaker diarization pipeline: principle,
  benchmark, and recipe},'' in \emph{Proc. INTERSPEECH 2023}, 2023, pp.
  1983--1987.

\bibitem{radford2023robust}
A.~Radford, J.~W. Kim, T.~Xu, G.~Brockman, C.~McLeavey, and I.~Sutskever,
  ``Robust speech recognition via large-scale weak supervision,'' in
  \emph{International Conference on Machine Learning}.\hskip 1em plus 0.5em
  minus 0.4em\relax PMLR, 2023, pp. 28\,492--28\,518.

\bibitem{valk2021slt}
J.~Valk and T.~Alum{\"a}e, ``{VoxLingua107}: a dataset for spoken language
  recognition,'' in \emph{Proc. IEEE SLT Workshop}, 2021.

\bibitem{vachhani23_interspeech}
B.~Vachhani, D.~Singh, and R.~Lawyer, ``{Multi-resolution Approach to
  Identification of Spoken Languages and To Improve Overall Language
  Diarization System Using Whisper Model},'' in \emph{Proc. INTERSPEECH 2023},
  2023, pp. 1993--1997.

\bibitem{landini2022bayesian}
F.~Landini, J.~Profant, M.~Diez, and L.~Burget, ``Bayesian {HMM} clustering of
  x-vector sequences (vbx) in speaker diarization: theory, implementation and
  analysis on standard tasks,'' \emph{Computer Speech \& Language}, vol.~71, p.
  101254, 2022.

\bibitem{google_ASR}
``Turn speech into text using google ai,''
  \url{https://cloud.google.com/speech-to-text?hl=en}, accessed: 2024-02-29.

\bibitem{watanabe2020chime}
S.~Watanabe, M.~Mandel, J.~Barker, E.~Vincent, A.~Arora, X.~Chang,
  S.~Khudanpur, V.~Manohar, D.~Povey, D.~Raj \emph{et~al.}, ``Chime-6
  challenge: Tackling multispeaker speech recognition for unsegmented
  recordings,'' \emph{arXiv preprint arXiv:2004.09249}, 2020.

\bibitem{kalda2024pixit}
J.~Kalda, R.~Marxer, T.~Alum{\"a}e, H.~Bredin \emph{et~al.}, ``Pixit: Joint
  training of speaker diarization and speech separation from real-world
  multi-speaker recordings,'' \emph{arXiv preprint arXiv:2403.02288}, 2024.

\bibitem{wang2023wespeaker}
H.~Wang, C.~Liang, S.~Wang, Z.~Chen, B.~Zhang, X.~Xiang, Y.~Deng, and Y.~Qian,
  ``Wespeaker: A research and production oriented speaker embedding learning
  toolkit,'' in \emph{ICASSP 2023-2023 IEEE International Conference on
  Acoustics, Speech and Signal Processing (ICASSP)}.\hskip 1em plus 0.5em minus
  0.4em\relax IEEE, 2023, pp. 1--5.

\bibitem{jia2021marblenet}
F.~Jia, S.~Majumdar, and B.~Ginsburg, ``Marblenet: Deep 1d time-channel
  separable convolutional neural network for voice activity detection,'' in
  \emph{ICASSP 2021-2021 IEEE International Conference on Acoustics, Speech and
  Signal Processing (ICASSP)}.\hskip 1em plus 0.5em minus 0.4em\relax IEEE,
  2021, pp. 6818--6822.

\bibitem{barrault2023seamless}
L.~Barrault, Y.-A. Chung, M.~C. Meglioli, D.~Dale, N.~Dong, M.~Duppenthaler,
  P.-A. Duquenne, B.~Ellis, H.~Elsahar, J.~Haaheim \emph{et~al.}, ``Seamless:
  Multilingual expressive and streaming speech translation,'' \emph{arXiv
  preprint arXiv:2312.05187}, 2023.

\bibitem{gusev2020deep}
A.~Gusev, V.~Volokhov, T.~Andzhukaev, S.~Novoselov, G.~Lavrentyeva, M.~Volkova,
  A.~Gazizullina, A.~Shulipa, A.~Gorlanov, A.~Avdeeva, A.~Ivanov, A.~Kozlov,
  T.~Pekhovsky, and Y.~Matveev, ``{Deep Speaker Embeddings for Far-Field
  Speaker Recognition on Short Utterances},'' in \emph{Proc. The Speaker and
  Language Recognition Workshop (Odyssey 2020)}, 2020, pp. 179--186.

\bibitem{lodagala2023ccc}
V.~S. Lodagala, S.~Ghosh, and S.~Umesh, ``Ccc-wav2vec 2.0: Clustering aided
  cross contrastive self-supervised learning of speech representations,'' in
  \emph{2022 IEEE Spoken Language Technology Workshop (SLT)}.\hskip 1em plus
  0.5em minus 0.4em\relax IEEE, 2023, pp. 1--8.

\end{thebibliography}

\end{document}